\documentclass[aps,twocolumn,superscriptaddress]{revtex4}
\usepackage{graphicx}

\newcommand{\ket}[1] {\left| #1 \right\rangle}

\begin{document}

\title{Landau-Zener transitions in a two-level system coupled to a finite-temperature harmonic oscillator}

\author{S. Ashhab}
\affiliation{Qatar Environment and Energy Research Institute, Qatar Foundation, Doha, Qatar}

\date{\today}


\begin{abstract}
We consider the Landau-Zener problem for a two level system (or qubit) when this system interacts with one harmonic oscillator mode that is initially set to a finite-temperature thermal equilibrium state. The oscillator could represent an external mode that is strongly coupled to the qubit, e.g.~an ionic oscillation mode in a molecule, or it could represent a prototypical uncontrolled environment. We analyze the qubit's occupation probabilities at the final time in a number of different regimes, varying the qubit and oscillator frequencies, their coupling strength and the temperature. In particular we find some surprising non-monotonic dependence on the coupling strength and temperature.
\end{abstract}


\maketitle

\section{Introduction}
\label{Sec:Introduction}

Landau-Zener (LZ) transitions occur when two energy levels cross, or more accurately experience an avoided crossing, as some external parameter is varied in time \cite{Landau,Zener,Stueckelberg,Majorana}. The system can then either stay in the same energy level that it occupied before the crossing, or it can undergo a transition to the other level. Such a universal phenomenon is ubiquitous and has applications in various areas of quantum physics. Among the new areas where the physics of LZ transitions can play an important role are adiabatic quantum computation (AQC) \cite{Farhi} and amplitude spectroscopy in nanoscale circuits \cite{Berns,Shevchenko,Johansson} and in nitrogen-vacancy centers in diamond \cite{Huang}. It could also play a role in the inter-molecular energy transfer in biological light-harvesting systems. In AQC, the parameters of a physical system (which can be called a quantum computer or annealer) are varied slowly such that the system transforms from an easy-to-prepare ground state into a ground state that contains the answer to a physical problem (or even a computational problem of non-physical nature). In biological light-harvesting systems, energy transfer between different parts of a molecule could be governed by molecular changes that act as driving fields for electronic motion.

The LZ problem in a closed system was solved soon after it was formulated over eighty years ago  \cite{Landau,Zener,Stueckelberg,Majorana}. Physical systems, however, invariably interact with a surrounding environment. There have been numerous studies on the LZ problem in the presence of an environment \cite{Kayanuma,Gefen,Ao,Shimshony,Nishino,Pokrovsky2003,Sarandy,Wubs,AshhabAQC,Lacour,Pokrovsky2007,Amin,Nalbach,Dodin,Xu,Haikka}, and some methods have produced accurate results in their regimes of validity. However, there is no method that is valid and computationally efficient for all parameter regimes. In particular, the different methods typically have underlying assumptions justifying the validity of their mathematical formulation based on physical arguments. For example, one could make the assumption of a very short correlation or memory time in the environment's degrees of freedom and use a Markovian approach. This approach would, however, break down when the environment's correlation time is not short compared to the LZ timescale, a situation that could occur when dealing with low-frequency noise.

Here we take a different approach. We numerically solve a rather simple physical problem that involves a single two-level system (to which we also refer as the qubit) coupled to a single harmonic oscillator. We can therefore be confident that our numerical calculations provide an accurate description of the problem as formulated. After obtaining the numerical results for the relatively simple problem, we comment on the physical significance of these results and how they could apply for a system where the single harmonic oscillator is replaced by an environment with a large number of degrees of freedom.

The remainder of this paper is organized as follows: In Sec.~\ref{Sec:Hamiltonian} we describe the basic setup and introduce the corresponding Hamiltonian. In Sec.~\ref{Sec:NumericalCalculations} we describe our numerical calculations. In Sec.~\ref{Sec:Results} we present the results of these calculations and discuss the interpretation of the results. Section \ref{Sec:Conclusion} contains some concluding remarks.

\section{Model system and Hamiltonian}
\label{Sec:Hamiltonian}

We consider the basic LZ problem where the system of interest possesses only two quantum states. As such, it can be described using the Pauli matrices $\hat{\sigma}_{\alpha}$ with $\alpha=x,y$ or $z$. We use the basis states defined by the relations $\hat{\sigma}_{z}\ket{\uparrow}=\ket{\uparrow}$ and $\hat{\sigma}_{z}\ket{\downarrow}=-\ket{\downarrow}$.

In an isolated system, the LZ Hamiltonian is given by
\begin{equation}
H = -\frac{vt}{2} \hat{\sigma}_{z}-\frac{\Delta}{2} \hat{\sigma}_{x},
\label{Eq:LZHamiltonianClosedSystem}
\end{equation}
where the time variable $t$ goes from $-\infty$ to $+\infty$, $v$ is the sweep rate and $\Delta$ is the minimum energy gap between the ground and excited states, which occurs at $t=0$. At large negative times the ground and excited states asymptotically coincide with the states $\ket{\downarrow}$ and $\ket{\uparrow}$, respectively. The roles of these states are reversed at large positive times. At $t=0$, the instantaneous ground and excited states are equal superpositions of the states $\ket{\uparrow}$ and  $\ket{\downarrow}$. The LZ formula, which for example gives the probability for a system prepared in its ground state at $t\rightarrow -\infty$ to end up in the excited state at $t\rightarrow \infty$, is given by $P_{\rm LZ} = \exp\{-\pi\Delta^2/(2v)\}$. In particular, for a slow sweep (i.e.~$v/\Delta^2\ll 1$), $P_{\rm LZ}\rightarrow 0$ and a system that is initially prepared in the ground state has a high probability of remaining in the ground state.

The LZ problem can be generalized in order to take into account the effects of an uncontrolled external environment. Early studies on this problem used somewhat ad hoc quantum master equations in order to incorporate dissipative processes in the dynamics \cite{Kayanuma}. Subsequent studies generally started with a specific model of the environment and derived approximate equations of motion for the system under certain approximations (see e.g.~Refs.~\cite{Ao,Wubs,Nalbach,Dodin}). Because of computational convenience and physical relevance, the environment is commonly modeled as a large set of harmonic oscillators, even if the microscopic details of the environment are not known. This approach has been applied successfully to the study of the LZ problem in a number of regimes. It is not possible, however, to obtain analytic results for this problem, and approximations that are valid for specific regimes are commonly made in order to numerically calculate the effect of the large number of harmonic oscillators on the LZ probability. The strong-coupling and low-temperature regimes are particularly challenging for these methods.

Here we take a different approach to studying the effects of the environment on the LZ problem. We consider an environment composed of a single harmonic oscillator. Clearly this simple model will not be able to capture all the effects that occur in a complex environment. However, the simplicity of the model allows us to have confidence in the results of standard numerical simulations. Rather than having to make assumptions concerning the behaviour of the system at the beginning of the calculation, the difficult task is then shifted to the step of interpreting the numerical results and identifying in these results patterns and tendencies that one can expect to apply for a large environment. It is also worth mentioning here that there can be cases where the largest environmental effects are caused by a single mode in the environment, in which case the results of this simple model become particularly relevant. Another advantage of treating such a minimal model is the fact that it allows us to discuss physical processes rather clearly.

We would like to note here that a related system, namely an LZ problem of a qubit coupled to a harmonic oscillator and an environment, was recently considered in Ref.~\cite{Zueco}. In that work, however, there is no minimum-gap term in the qubit's Hamiltonian, and the avoided crossings arise as a result of the coupling between the qubit and the oscillator, rendering the system qualitatively different from the one that we consider in this paper.

\begin{figure}[h]
\includegraphics[width=8.0cm]{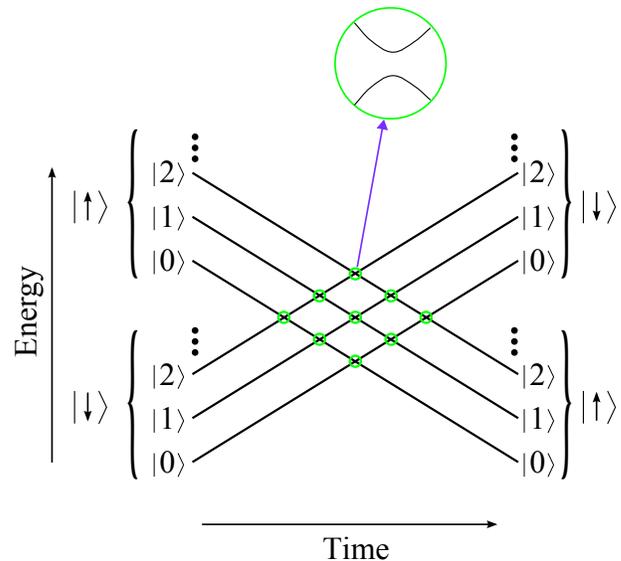}
\caption{(color online) Energy level diagram of a coupled qubit-oscillator system with the qubit bias conditions varied according to the LZ protocol.}
\label{Fig:EnergyLevelDiagram}
\end{figure}

The Hamiltonian of the LZ problem with a single-mode environment and linear coupling is given by:
\begin{equation}
H = -\frac{vt}{2} \hat{\sigma}_{z} - \frac{\Delta}{2} \hat{\sigma}_{x} + \hbar \omega \hat{a}^{\dagger} \hat{a} + g \sigma_z \otimes \left( \hat{a} + \hat{a}^{\dagger} \right),
\label{Eq:LZHamiltonianSingleEnvironmentMode}
\end{equation}
where $\omega$ is the characteristic frequency of the harmonic oscillator, $\hat{a}$ and $\hat{a}^{\dagger}$ are, respectively, the oscillator's annihilation and creation operators, and $g$ is the qubit-oscillator coupling strength. The energy level diagram of this problem is illustrated in Fig.~\ref{Fig:EnergyLevelDiagram}.

We are interested in particular in the case of slow, nearly adiabatic passage. This case corresponds to the desired condition for obtaining a high transfer probability in adiabatic passage protocols; it is also the relevant regime for maximizing the success probability in an adiabatic quantum computation.

\section{Numerical calculations}
\label{Sec:NumericalCalculations}

We numerically solve the time-dependent Schr\"odinger (or Liouville-von Neumann) equation using the Hamiltonian given in Eq.~(\ref{Eq:LZHamiltonianSingleEnvironmentMode}). In these calculations we set the sweep rate $v$ to the value that gives $P_{\rm LZ}=0.1$ (i.e., starting from the ground state, the two-level system ends up in the ground state with 90\% probability). In other words, we choose a sweep rate that is close to the adiabatic limit in the absence of the coupling to the oscillator. We take three different values for the oscillator frequency: $\omega/\Delta=0.2$ (low-frequency oscillator), 1 (intermediate regime), and 5 (high-frequency oscillator). We vary the coupling strength from $g/\Delta=0$ to $g/\Delta=2$, and we vary the temperature $T$ from $k_BT/\Delta=0$ to $k_BT/\Delta=5$, where $k_B$ is the Boltzmann constant.

In order to incorporate the finite temperature into the calculation, the simulations are started in thermal equilibrium at a large negative value for the time variable. In this limit, the qubit and resonator are effectively decoupled from each other, except for simple mean-field shifts that they induce on each other. Furthermore, the qubit's energy splitting is very large in the limit $t\rightarrow -\infty$. As a result, the qubit starts initially in its ground state $\ket{\downarrow}$. The harmonic oscillator starts in a mixed thermal state according to the Boltzmann probability distribution with an average number of quanta $k_BT/(\hbar\omega)$ for high temperatures (Note that the Boltzmann probability distribution extends up to several times this value). This estimate provides a minimum number of basis states that need to be included in the simulations, and it also sets a limit to the highest temperatures that can be reached in simulations with a given size of the Hilbert space. In particular for the lowest oscillator frequency and highest temperature that we consider, we use a Hilbert space with 1000 basis states. Note that the initial state of the oscillator is the only part of the calculation where the finite temperature of the environment enters the calculation.

After setting the initial state according to the Boltzmann distribution, we evolve the density matrix of the combined system in time according to the Schr\"odinger equation. Note that this evolution is unitary, which is the reason why we can say that, in contrast to most other methods, we do not make any approximations or assumptions concerning the internal dynamics of the environment. The evolution is stopped at a sufficiently large and positive value of the time variable, such that further evolution would not have any noticeable effect on the occupation probabilities of the different states. At this final time, we examine the occupation probabilities of the different quantum states, from which we can easily calculate the probability that the qubit remains in its ground state.

\begin{figure}[h]
\includegraphics[width=7.5cm]{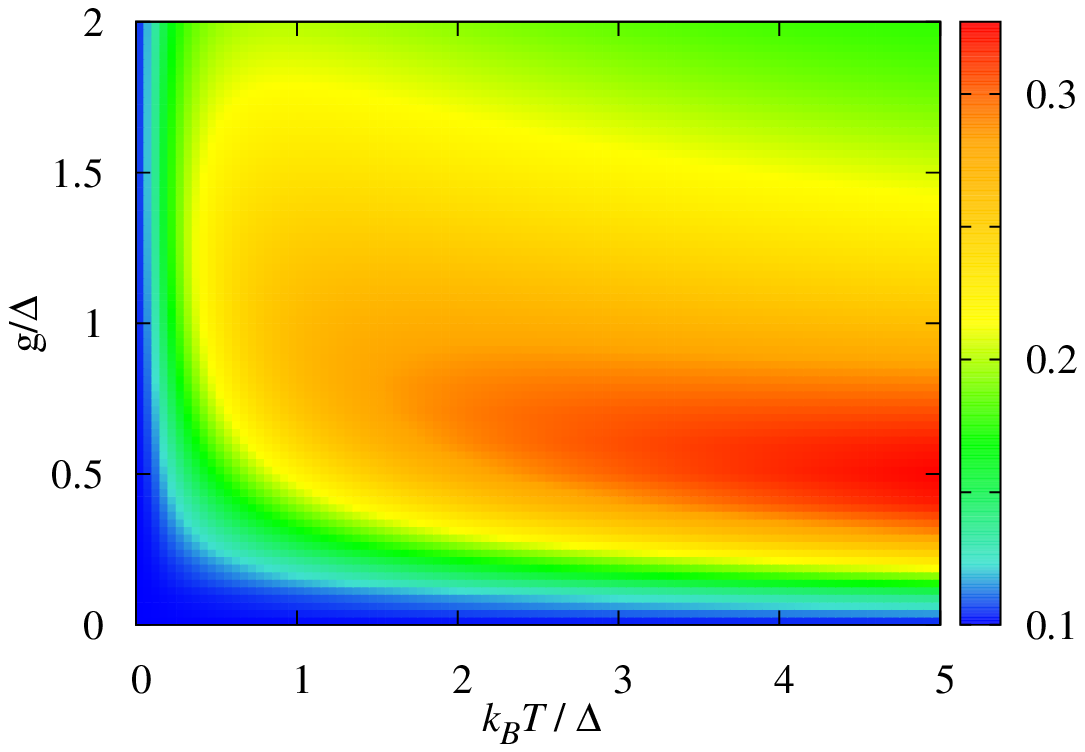}
\includegraphics[width=7.0cm]{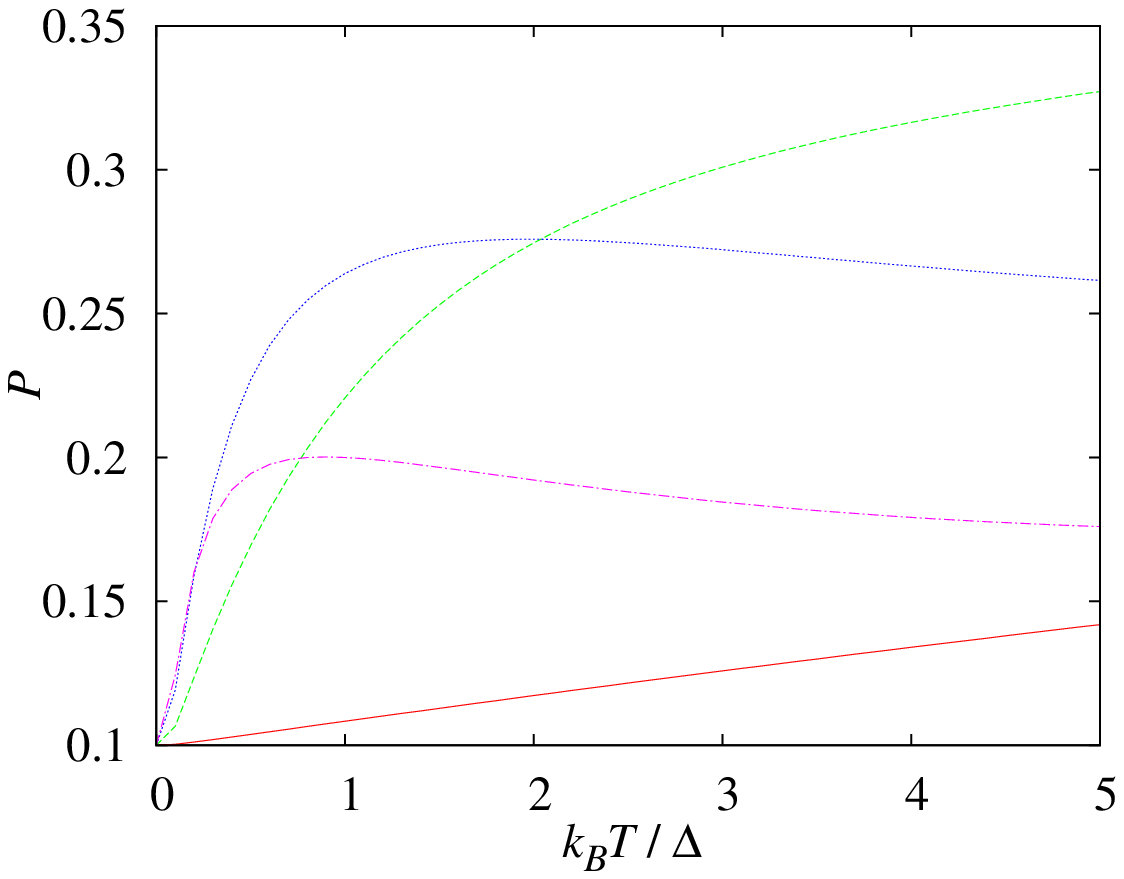}
\includegraphics[width=7.0cm]{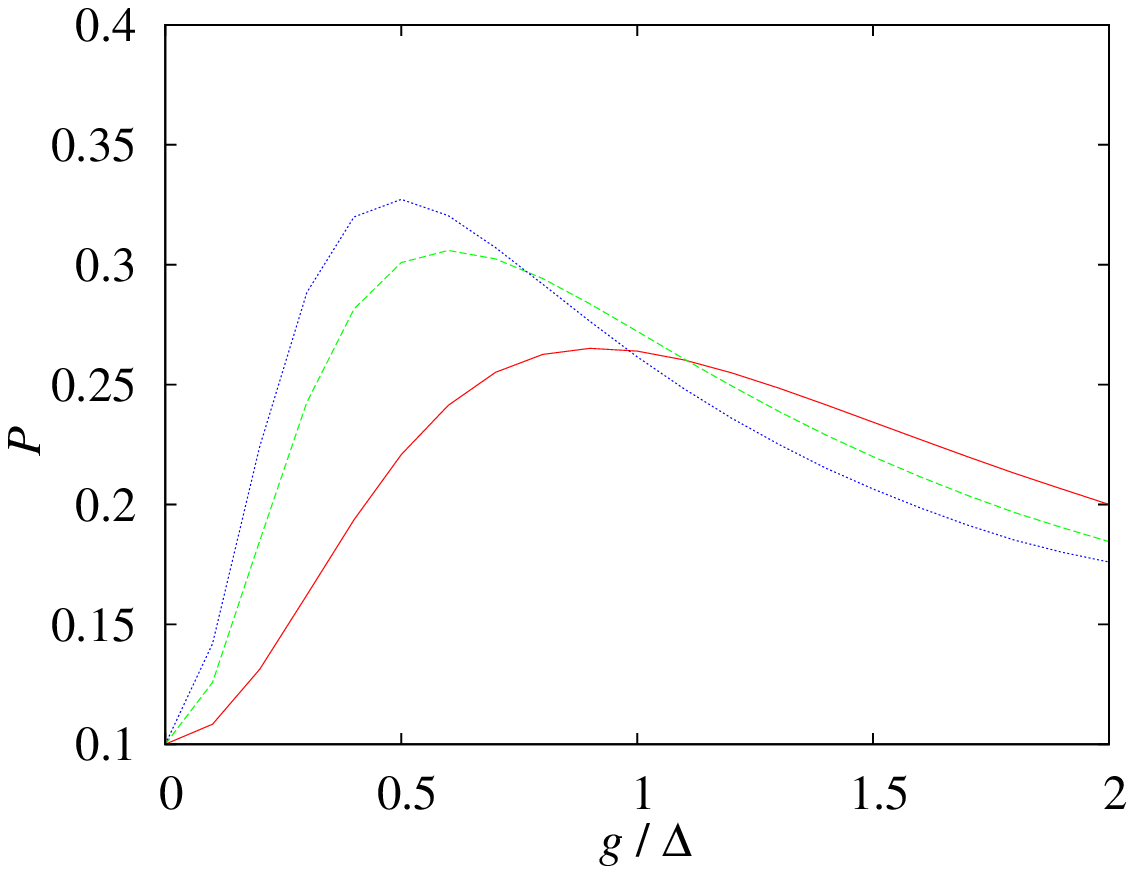}
\caption{(color online) Top: Qubit's final excited-state probability $P$ as a function of temperature $k_BT$ and coupling strength $g$, both measured relative to the qubit's minimum gap $\Delta$. Middle: $P$ as a function of $k_BT/\Delta$ for four different values of $g/\Delta$: 0.1 (red solid line), 0.3 (green dashed line), 1 (blue dotted line) and 2 (magenta dash-dotted line). Bottom: $P$ as a function of $g/\Delta$ for three different values of $k_BT/\Delta$: 1 (red solid line), 3 (green dashed line), and 5 (blue dotted line). In all the panels, the harmonic oscillator frequency is $\hbar\omega/\Delta=0.2$. The sweep rate is chosen such that $P_{\rm LZ}=0.1$, and this value is the baseline for all of the results plotted in this figure.}
\label{Fig:ExcitationProbability02}
\end{figure}

\begin{figure}[h]
\includegraphics[width=7.5cm]{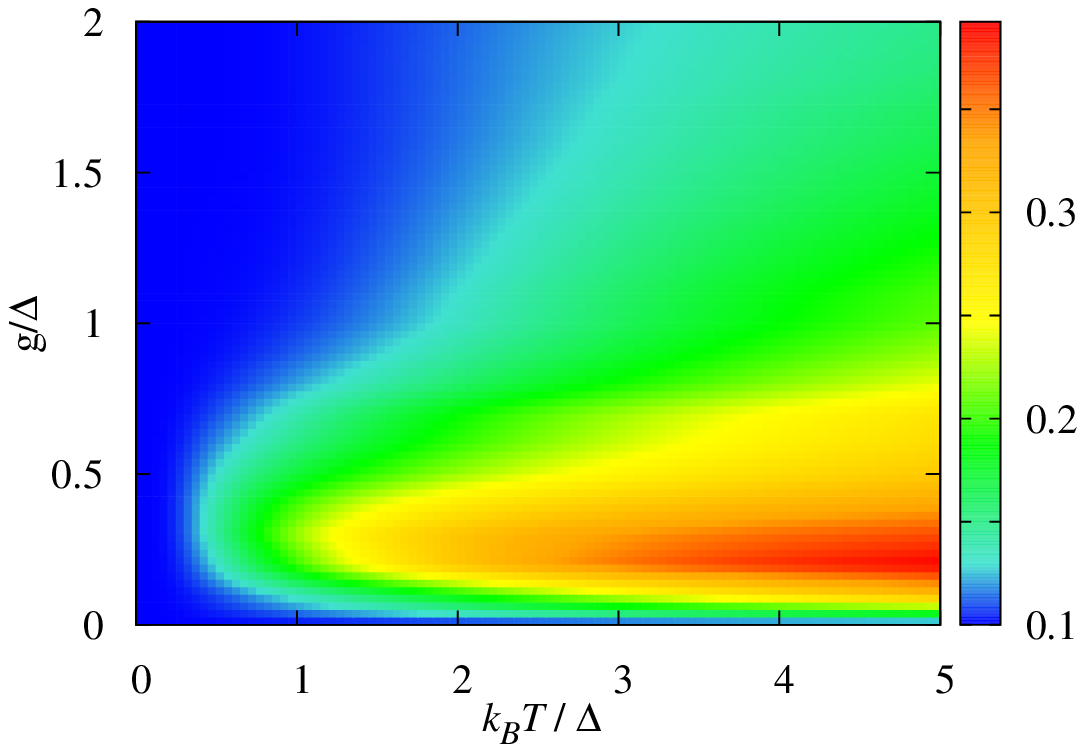}
\includegraphics[width=7.0cm]{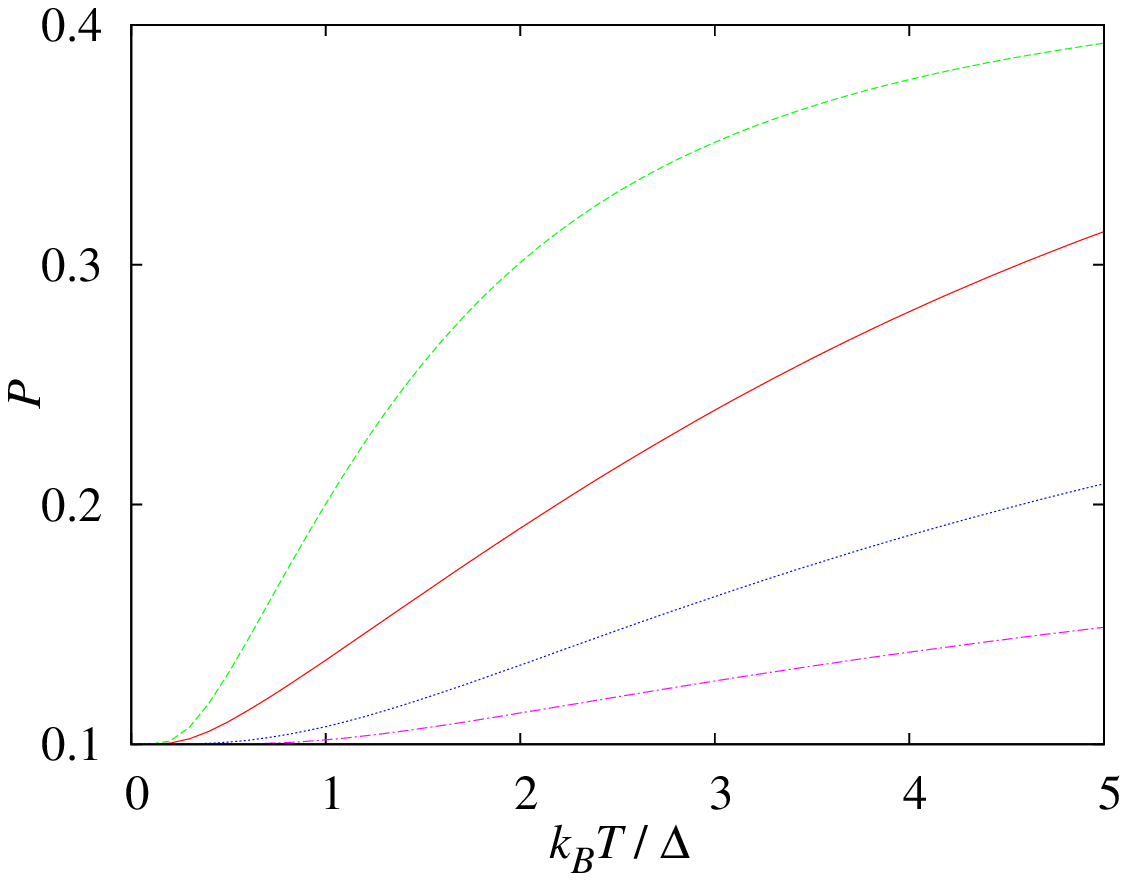}
\includegraphics[width=7.0cm]{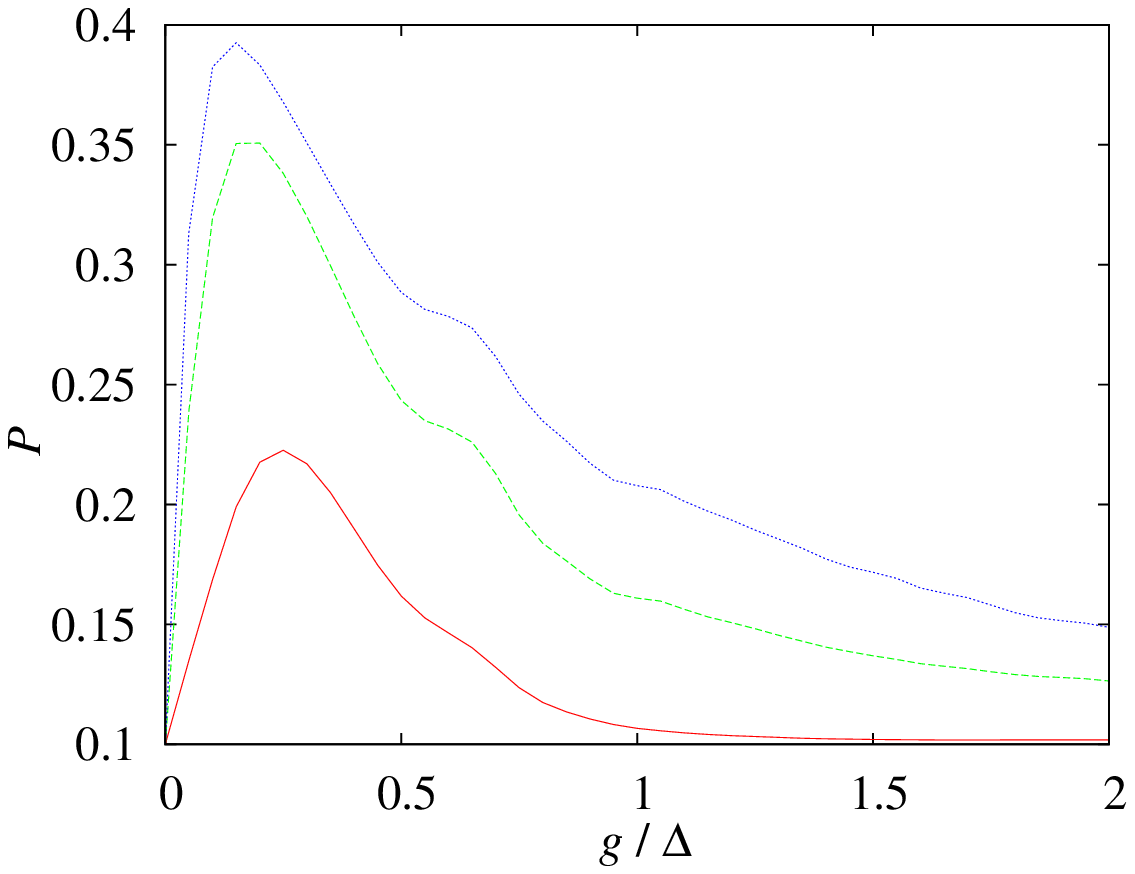}
\caption{(color online) Same as in Fig.~\ref{Fig:ExcitationProbability02} but for $\hbar\omega/\Delta=1$.}
\label{Fig:ExcitationProbability10}
\end{figure}

\begin{figure}[h]
\includegraphics[width=7.5cm]{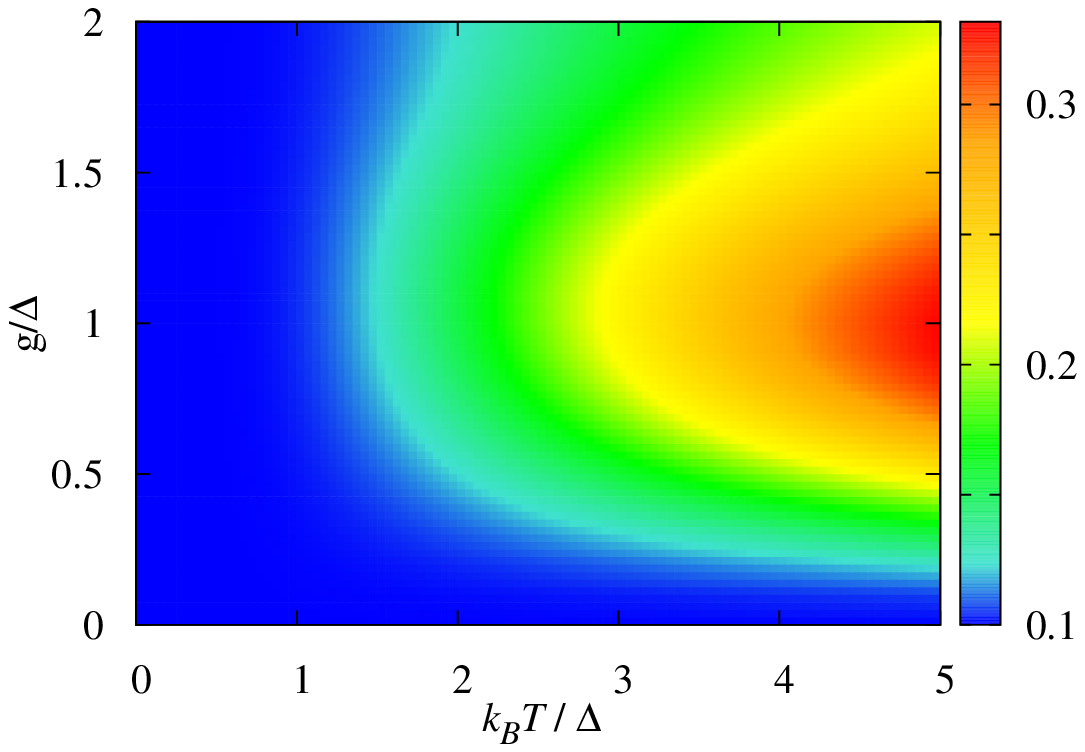}
\includegraphics[width=7.0cm]{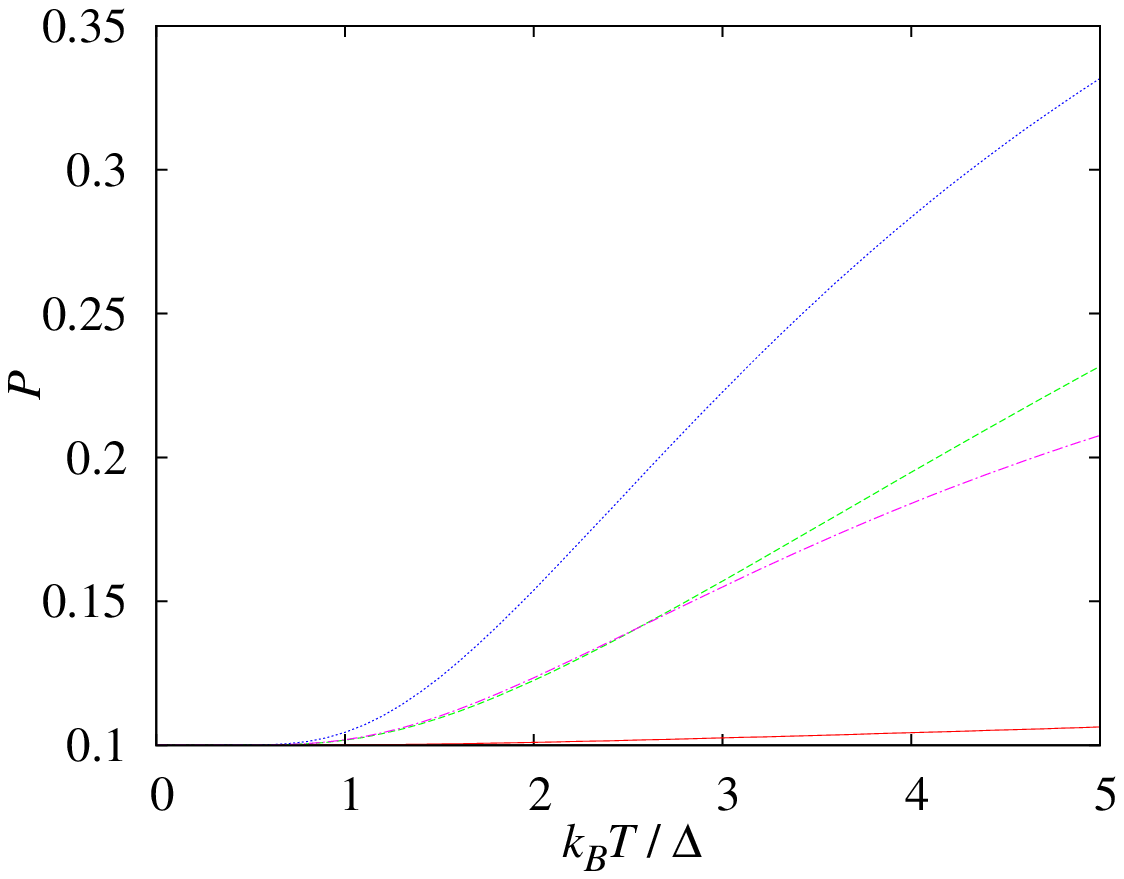}
\includegraphics[width=7.0cm]{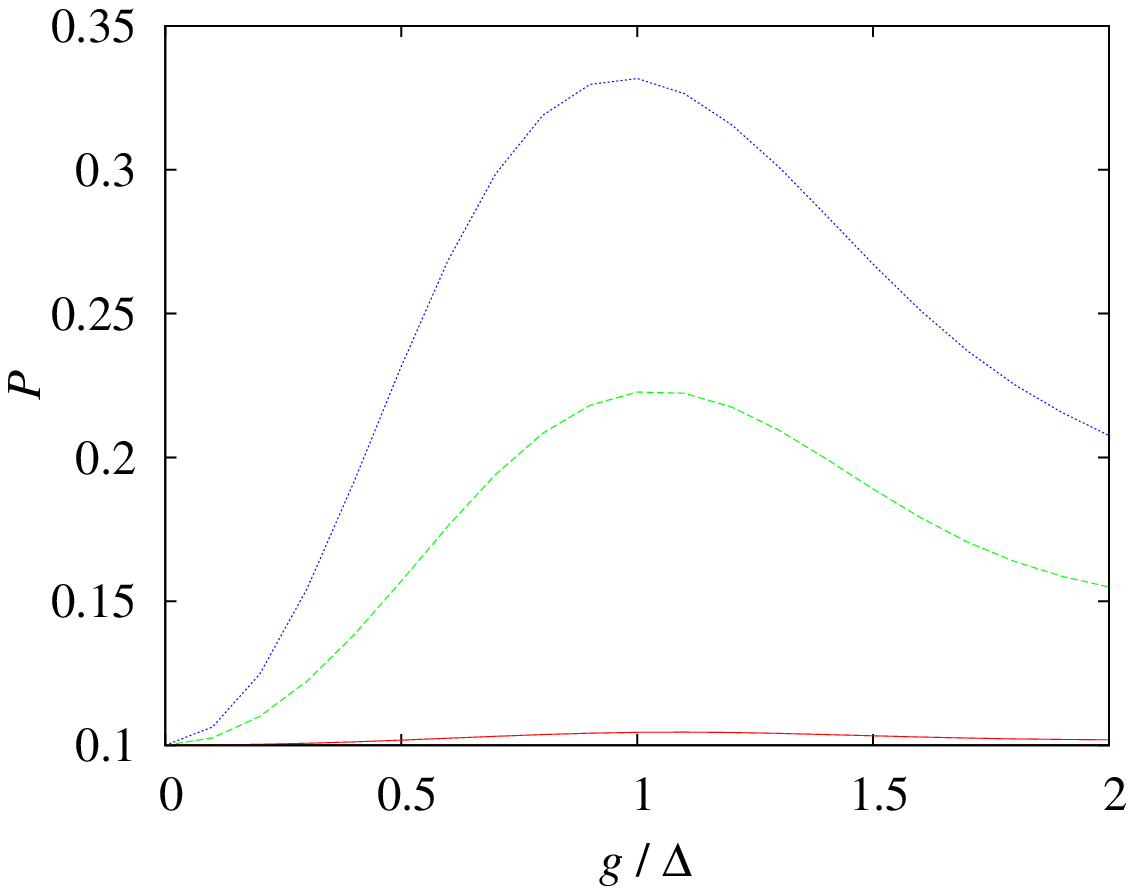}
\caption{(color online) Same as in Fig.~\ref{Fig:ExcitationProbability02} but for $\hbar\omega/\Delta=5$.}
\label{Fig:ExcitationProbability50}
\end{figure}

\section{Results}
\label{Sec:Results}

The probability for the qubit to end up in the excited state at the final time as a function of temperature and coupling strength is plotted in Figs.~\ref{Fig:ExcitationProbability02}-\ref{Fig:ExcitationProbability50}. As expected from known results \cite{Wubs}, the final excited-state occupation probability $P$ remains equal to 0.1 whenever the temperature or the coupling strength is equal to zero. Otherwise, the coupling to the oscillator causes this probability to increase. A common, and somewhat surprising, trend for all values of $\hbar\omega/\Delta$ is the non-monotonic dependence on the coupling strength $g$. As the coupling strength is increased from zero to finite but small values, $P$ increases. But when the coupling strength is increased further, $P$ starts decreasing. Based on the results that are plotted in Figs.~\ref{Fig:ExcitationProbability02}-\ref{Fig:ExcitationProbability50}, one can expect that in the limit of large $g/\Delta$ (and assuming not-very-large values of $k_BT/\Delta$) the excited-state occupation probability will go back to its value in the uncoupled case, i.e.~$P=0.1$. This phenomenon is probably a manifestation of the superradiance-like behaviour in a strongly coupled qubit-oscillator system \cite{AshhabSuperradiance}. In the superradiant regime (i.e.~the strong-coupling regime), the ground state is highly entangled exactly at the symmetry point (which corresponds to the bias conditions at $t=0$ in the LZ problem), but even small deviations from the symmetry point can lead to an effective decoupling between the qubit and resonator with the exception of some state-dependent mean-field shifts. Indeed the maximum values of $P$ reached in Figs.~\ref{Fig:ExcitationProbability10} and \ref{Fig:ExcitationProbability50} occur at coupling strength values that are comparable to the expression for the uncorrelated-to-correlated crossover value, namely $g\sim\hbar\omega$ (and we have verified that the near-linear increase in peak location as a function of oscillator frequency continues up to $\hbar\omega/\Delta=20$). This relation does not apply in the case $\hbar\omega/\Delta=0.2$, shown in Fig.~\ref{Fig:ExcitationProbability02}. In this case, the peak occurs when the coupling strength $g$ is comparable to the minimum gap $\Delta$. It is in fact quite surprising that the excitation peak in the case $\hbar\omega/\Delta=0.2$ occurs at a higher coupling strength than that obtained in the case $\hbar\omega/\Delta=1$. In order to investigate this point further, we tried values close to $\hbar\omega/\Delta=1$ and found that this value gives a minimum in the peak location (i.e.~the peak in $P$ when plotted as a function of $g/\Delta$).

Another feature worth noting is the temperature dependence of $P$ close to zero temperature. As can be seen clearly in Figs.~\ref{Fig:ExcitationProbability10} and \ref{Fig:ExcitationProbability50}, the initial increase in $P$ with temperature is very slow, indicating that it probably follows an exponential function that corresponds to the probability of populating the excited states in the harmonic oscillator (and the same dependence is probably present but difficult to see because of the scale of the $x$ axis in Fig.~\ref{Fig:ExcitationProbability02}). After this initial slow rise, and in particular when $k_BT \gtrsim \hbar\omega$, we see a steady rise that in the case of Fig.~\ref{Fig:ExcitationProbability02} can be approximated as a linear increase in $P$ with increasing $T$. Importantly, the slope of this increase can be quite large for intermediate $g$ values. From the results shown in Figs.~\ref{Fig:ExcitationProbability02}-\ref{Fig:ExcitationProbability50}, we find that the maximum slope $[dP/d(k_BT/\Delta)]_{\rm max}=0.18\times(\hbar\omega/\Delta)^{-0.57}$, and results for other parameter values extending up to $\hbar\omega/\Delta=20$ follow this dependence. The implication of this result can be seen clearly in the middle panel of Fig.~\ref{Fig:ExcitationProbability02}: even when the temperature is substantially smaller than the qubit's minimum gap $\Delta$, the initial excitation of the low-frequency oscillator (stemming from the finite temperature) can cause a large increase in the qubit's final excited-state probability. This result is in contrast with the exact result of Ref.~\cite{Wubs} stating that at zero temperature the qubit's final excited-state probability is given by $P_{\rm LZ}$ regardless of the value of $g$. The typical temperature scale at which deviations from the LZ formula occur can therefore be much lower than $\Delta/k_B$. This result is relevant for adiabatic quantum computing, because it contradicts the expectation that having a minimum gap that is large compared to the temperature might provide automatic protection for the ground state population against thermal excitation. Another point worth noting here is that when $\hbar\omega<\Delta$, there is no point in time where the qubit and oscillator are resonant with each other, yet the initial thermal excitation of the oscillator can result in exciting the qubit at the final time. The excitations in the oscillator are in some sense up-converted into excitations in the qubit as a result of the sweep through the avoided crossing.

\begin{figure}[h]
\includegraphics[width=7.0cm]{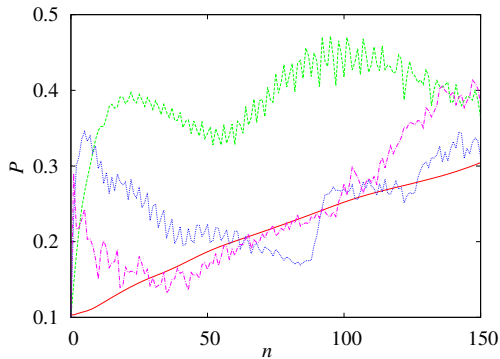}
\caption{(color online) The final excited state probability $P$ as a function of the number of excitation quanta $n$ present in the initial state of the oscillator. Here we take $\hbar\omega/\Delta=0.2$. The different lines correspond to different values of the coupling strength: $g/\Delta=0.1$ (red solid line), 0.5 (green dashed line), 1 (blue dotted line) and 2 (magenta dash-dotted line).}
\label{Fig:ExcitationProbabilityAsFunctionOfInitialOscillatorExcitationNumber}
\end{figure}

We can also see in Fig.~\ref{Fig:ExcitationProbability02} that for $g/\Delta\gtrsim 1$ the temperature dependence is non-monotonic. In particular, for low temperatures we obtain the intuitively expected increase in excitation probability with increasing temperature, but this trend reverses for higher temperatures. In order to investigate this feature further, we calculate the qubit's final excited-state probability as a function of the number $n$ of excitation quanta present in the initial state of the oscillator (Note that this calculation differs from the ones described above in that here we do not use the Boltzmann distribution for the oscillator's initial state). The results are plotted in Fig.~\ref{Fig:ExcitationProbabilityAsFunctionOfInitialOscillatorExcitationNumber}. These results explain the non-monotonic dependence on temperature. For intermediate values of $g/\Delta$ (e.g.~for $g/\Delta=1$), there is a peak at a small but finite excitation number followed by a steady decrease. As the temperature is increased from zero, the qubit's final excited-state probability samples the probabilities for increasingly high excitation numbers, and a peak at intermediate values of temperature is obtained. Note that for large excitation numbers, the increase in $P$ as a function of $n$ resumes, and this increase will also be reflected in the temperature dependence.

We note in this context that recent theoretical studies \cite{Nalbach,Dodin} have reported non-monotonic dependence of the excitation probability as a function of the sweep rate $v$. However, that dependence was generally oscillatory, and we suspect that it has a different origin from the behaviour obtained in the present study. We expect that similar oscillatory behaviour would be obtained if we varied $v$ in our calculations. As mentioned in Sec.~\ref{Sec:Hamiltonian}, however, here we are mainly interested in the almost-adiabatic regime, and we have therefore not analyzed the $v$ dependence in our calculations.

\begin{figure}[h]
\includegraphics[width=7.5cm]{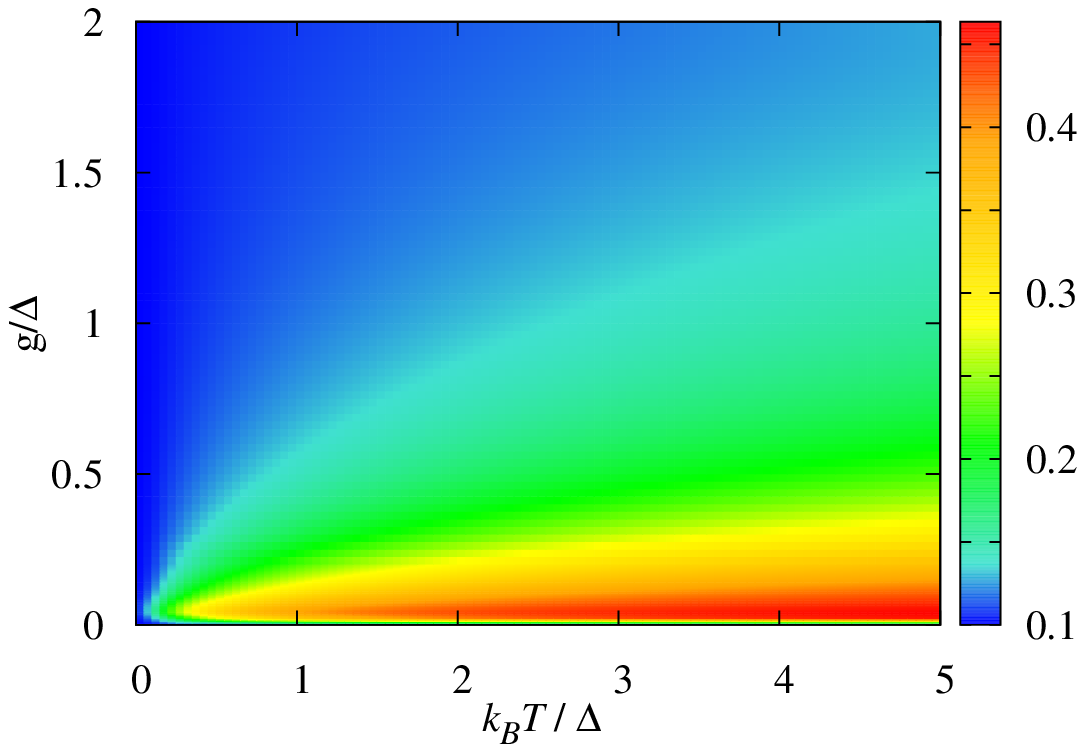}
\includegraphics[width=7.5cm]{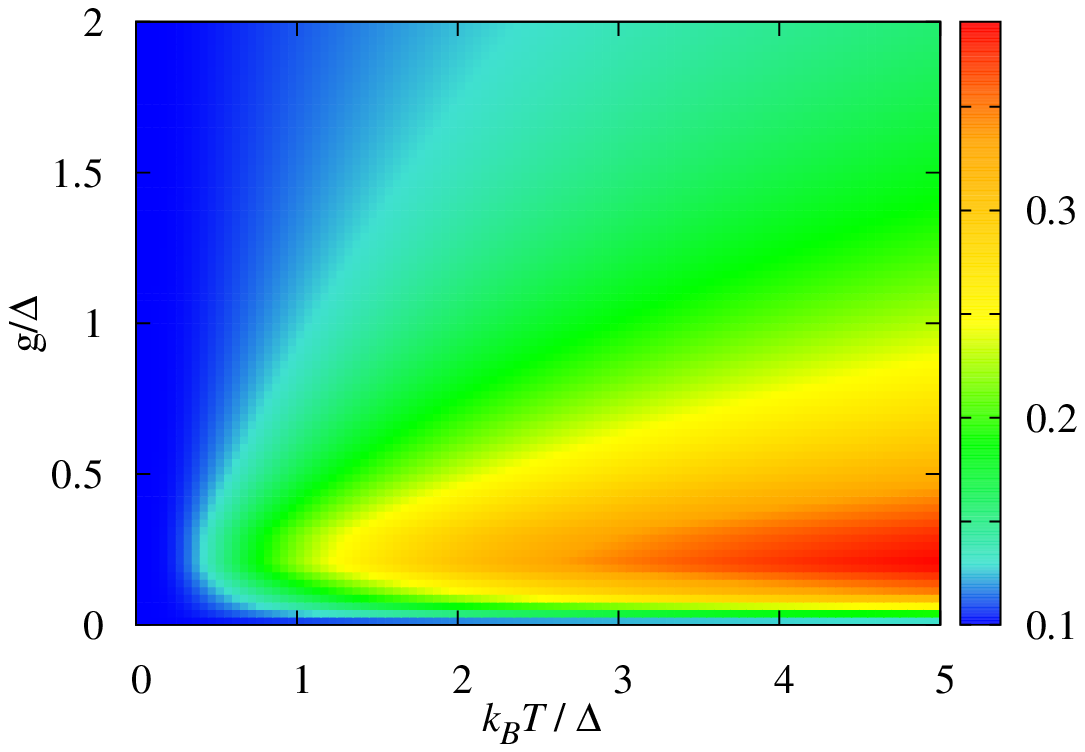}
\includegraphics[width=7.5cm]{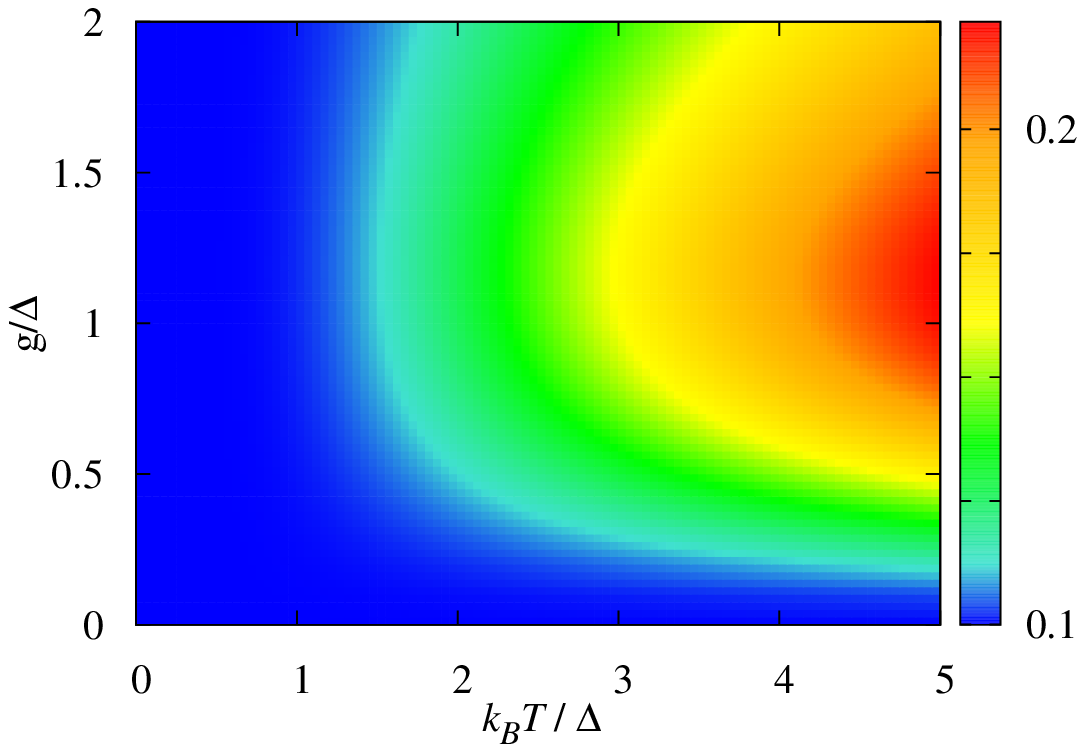}
\caption{(color online) Qubit's final excited state probability $P$ obtained from the semiclassical calculation as a function of temperature $k_BT$ and coupling strength $g$, both measured relative to the minimum qubit gap $\Delta$. The different panels correspond to different values of the harmonic oscillator frequency: $\hbar\omega/\Delta=0.2$ (top), 1 (middle) and 5 (bottom).}
\label{Fig:ExcitationProbabilityFromIncoherentCalculation}
\end{figure}

In addition to solving the Schr\"odinger equation, we have performed semiclassical calculations where we assume that there is no quantum coherence between the different LZ processes. (Note here that when we replace the isolated qubit with the coupled qubit-oscillator system the single avoided crossing is replaced by a complex network of avoided crossings.) Under this approximation, we only need to calculate the occupation probabilities of the different states, and these probabilities change (according to the LZ formula) only at the points of avoided crossing. This approach greatly simplifies the numerical calculations because the locations and gaps for the different avoided crossings can be determined easily (see e.g.~Fig.~\ref{Fig:EnergyLevelDiagram}). The results are shown in Fig.~\ref{Fig:ExcitationProbabilityFromIncoherentCalculation}. The results of this calculation agree generally well with those obtained by solving the Schr\"odinger equation when $\hbar\omega/\Delta=1$. For $\hbar\omega/\Delta =5$, the semiclassical calculation consistently underestimates the excited-state probability, but the overall dependence on temperature and coupling strength is remarkably similar to that shown in Fig.~\ref{Fig:ExcitationProbability50}. We should note that higher values of $\hbar\omega$ (not shown) exhibit more pronounced deviations, with side peaks appearing in the dependence of $P$ on $g/\Delta$. The most striking deviation from the results of the fully quantum calculation is seen in the case $\hbar\omega/\Delta=0.2$ (i.e.~the case of a low-frequency oscillator). In the semiclassical calculation, there is a rather high peak at a small value of the coupling strength (and sufficiently high temperatures), and the excited-state probability starts decreasing when the coupling strength $g$ becomes larger than $\hbar\omega$. In the fully quantum calculation, however, the peak is located at a much higher value, somewhere between 0.5 and 1 depending on the temperature.

The fact that the semiclassical calculation generally gives results different from those given by the fully quantum calculation is an indication that quantum coherence and interference between multiple LZ processes play a role in determining the final occupation probabilities. In this context we note that the avoided crossings occur at instances separated by time intervals $\tau_{\rm separation}=\hbar\omega/v$ (with an infinite number of avoided crossings occurring simultaneously at each one of these instances), and the time duration over which an LZ mixing process occurs (in the almost-adiabatic regime) is given by $\tau_{\rm LZ}\sim\Delta/v$ \cite{Shevchenko,MinimumGapFootnote}. The ratio between these two time scales is then given by $\tau_{\rm separation}/\tau_{\rm LZ}\sim\hbar\omega/\Delta$. In other words, when $\hbar\omega/\Delta$ is small the different LZ processes will overlap in time, and it is not too surprising that the semiclassical calculation gives incorrect predictions in this case. It is somewhat surprising, however, that when $\hbar\omega/\Delta=1$ the two calculations agree quite well and then in the regime $\hbar\omega/\Delta>1$ the effect of quantum interference between the LZ processes can again be seen in the final occupation probabilities.

We now take another look at our results presented above from the point of view of how they might apply in the case of a large environment containing a large number of degrees of freedom with no single dominant environmental mode. Note here that the coupling between the qubit and the environment can in principle be strong, even if the coupling to each individual mode in the environment is weak. We first consider our results in the regime of strong qubit-environment coupling. We have found that strong coupling to a single mode results in a reduced effect of that mode on the final occupation probabilities. It is unlikely that this result will apply to the case where the qubit is coupled strongly to an uncontrolled environment containing a large number of independent modes with the coupling to each individual mode in the environment being weak. The weakening of the environmental effects with increased coupling strength in the case of a single mode is most likely related to the energy level structure and the possible paths that the system can follow while it traverses the network of avoided crossings. The energy level structure and the possible paths are vastly different when the strong coupling to the environment is caused by the large number of modes in the environment. It would be more plausible that in this case one can make statements concerning large environments using the following approach: focus on the small $g/\Delta$ region of the results discussed above, take the contributions of the individual environment modes and add up these small contributions. In this case an increase in coupling strength would result in an increase in the excited state probability, as would be intuitively expected. We therefore expect that the result of non-monotonic behaviour with increasing coupling strength should be thought of as a result pertaining to the case with a single dominant mode in the environment. Another area where we can try to extract from our results statements concerning a large environment occurs in the regime of low temperatures, which can be particularly relevant in the context of AQC. As a side note, we mention here that one of the central questions in the field of AQC is the scaling of the minimum gap with system size. It is known that the minimum gap decreases with increasing system size, and there are ongoing studies on the exact scaling law. This minimum-gap scaling is typically discussed in relation to the time needed to ensure adiabatic evolution of the quantum annealer, and the minimum running time is calculated based on the well-known LZ formula given in Sec.~\ref{Sec:Hamiltonian}. An independent question is the resistance of the AQC success probability to environmental noise. The facts that at finite temperatures the excitation probability increases above the base value $P_{\rm LZ}$ and that the excitation probability can be substantially larger than $P_{\rm LZ}$ even at temperatures much lower than the minimum gap mean that the coupling to the low-frequency modes in the environment needs to be considered with extra care in the low-temperature regime. In a previous work \cite{AshhabAQC}, we discussed the scaling of the noise amplitude with system size, with the main message being that the noise amplitude increases with increasing system size. The present work complements our earlier work in that it provides a quantitative analysis of the effect of the environment on a system driven using an adiabatic passage protocol, as is the case in AQC.

\section{Conclusion}
\label{Sec:Conclusion}

We have investigated the problem of a two-level system undergoing an LZ passage through an avoided crossing while it interacts with a finite-temperature harmonic oscillator. We have found a number of counter-intuitive results, including non-monotonic dependence of the final-time excitation probability as a function of temperature or qubit-oscillator coupling strength. We have provided physical explanations for these phenomena. The physical mechanisms at play include modifications to the avoided crossing structure related to the formation of highly correlated energy eigenstates as well as quantum coherence between multiple LZ processes.

Our original motivation for analyzing a system with a single qubit and a single additional degree of freedom was to use the obtained results in order to make statements relevant for a large environment, and we have indeed attempted to make such an extrapolation of results. We emphasize, however, that our results are of interest even in relation to the single-oscillator case, both because they pertain to a model system that allows a clear discussion of the physical mechanisms involved and because certain systems in nature are accurately described by the model of a single qubit coupled to a single oscillator. In other words, in addition to the general principles that we have deduced concerning general environments, our results can have direct applicability to qubit-oscillator systems such as cavity electrodynamics or some molecular systems.


\begin{thebibliography}{99}
\bibitem{MinimumGapFootnote} Note that most of the avoided crossing gaps are much smaller than $\Delta$, and this estimate is therefore a conservative one.
\end{thebibliography}
\end{document}